\documentclass[aps,prl,twocolumn,superscriptaddress]{revtex4}

\usepackage{mathrsfs,amssymb,graphics,subfigure,threeparttable}
\usepackage{grap hicx}
\usepackage{amssymb}
\usepackage{enumerate}
\usepackage{amsmath}
\usepackage{fancyhdr}
\usepackage{bm}
\usepackage[squaren]{SIunits}
\usepackage{xcolor}
\usepackage{changes}

\begin{document}

\title{Generation and acceleration of high brightness electrons beams bunched at X-ray wavelengths using plasma-based acceleration
}





\author{Xinlu Xu}
\email[]{xuxinlu@slac.stanford.edu}
\affiliation{SLAC National Accelerator Laboratory, Menlo Park, CA 94025}
\author{Fei Li}
\affiliation{Department of Electrical Engineering, University of California, Los Angeles, California 90095, USA}
\author{Frank S. Tsung}
\affiliation{Department of Physics and Astronomy, University of California Los Angeles, Los Angeles, CA 90095, USA}
\author{Kyle Miller}
\affiliation{Department of Electrical Engineering, University of California, Los Angeles, California 90095, USA}
\author{Vitaly Yakimenko}
\affiliation{SLAC National Accelerator Laboratory, Menlo Park, CA 94025}
\author{Mark J. Hogan}
\affiliation{SLAC National Accelerator Laboratory, Menlo Park, CA 94025}
\author{Chan Joshi}
\affiliation{Department of Electrical Engineering, University of California, Los Angeles, California 90095, USA}
\author{Warren B. Mori}
\affiliation{Department of Electrical Engineering, University of California, Los Angeles, California 90095, USA}
\affiliation{Department of Physics and Astronomy, University of California Los Angeles, Los Angeles, CA 90095, USA}


\date{\today}

\begin{abstract} 
We show using particle-in-cell (PIC) simulations and theoretical analysis that a high-quality electron beam whose density is modulated at angstrom scales can be generated directly using density downramp injection in a periodically modulated density in nonlinear plasma wave wakefields. The density modulation turns on and off the injection of electrons at the period of the modulation. Due to the unique longitudinal mapping between the electrons' initial positions and their final trapped positions inside the wake, this results in an electron beam with density modulation at a wavelength orders of magnitude shorter than the plasma density modulation. The ponderomotive force of two counter propagating lasers of the same frequency can generate a density modulation at half the laser wavelength. Assuming a laser wavelength of $0.8\micro\meter$, fully self-consistent OSIRIS PIC simulations show that this scheme can generate high quality beams modulated at wavelengths between 10s and 100 angstroms. Such beams could produce fully coherent, stable, hundreds of GW X-rays by going through a resonant undulator. 
\end{abstract}

\pacs{}

\maketitle


X-rays - here defined as electromagnetic radiation with wavelength ranging from $\sim$100 to $\sim$1 Angstroms - have made great contributions in modern science, industry and medicine since they were discovered in 1895 \cite{rontgen1896new}. X-ray free-electron-lasers (FELs)  can deliver the most powerful beam of directional and coherent X-rays at tunable wavelengths \cite{pellegrini2016physics, kim2017synchrotron}. The XFEL process relies on an instability that arises when a bright electron beam propagates in an undulator that wiggles the electrons transversely. Current X-FELs operate in the self-amplified spontaneous emission (SASE) mode which produces X-ray photons with a limited longitudinal coherence arising from the initial chaotic, spontaneous synchrotron radiation. A seed laser pulse co-propagating with the electrons can induce longitudinal phase correlation across the whole beam, thus improving the temporal coherence, stability of the pulse energy and narrowing the spectrum. Due to the lack of an intense, coherent seed at the X-ray wavelength, harmonic seeding schemes, such as cascaded high-gain harmonic generation (HGHG) \cite{yu1991generation,yu1997high,yu2000high,allaria2012highly, allaria2013two} and echo-enabled harmonic generation (EEHG) \cite{stupakov2009using, xiang2009echo, xiang2010demonstration, zhao2012first, xiang2012evidence, hemsing2016echo, ribivc2019coherent}, have been studied extensively. In these schemes, one or two laser pulses, multiple undulators and magnetic chicanes are used to convert the wavelength of the electron density modulation from the seed laser wavelength ($\hbar\omega \sim 3$ eV) to its high harmonics with a harmonics number $h \leq 101$ \cite{allaria2013two, ribivc2019coherent} ($\hbar\omega \sim 300$ eV). 

Due to its ability to sustain $ \giga\volt/\centi\meter$ acceleration gradient, plasma-based acceleration (PBA) driven by either an intense laser pulse or a high current charged particle beam \cite{PhysRevLett.43.267, chen1985acceleration, joshi2006plasma} can accelerate electrons to GeV-level energies in only a few centimeters \cite{blumenfeld2007energy, litos2014high,PhysRevLett.122.084801, adli2018acceleration}. Furthermore, numerical experiments show that high quality (i.e., high brightness, low energy spread) beams suitable for driving an XFEL in the SASE mode can be generated from PBA \cite{xu2017high, PhysRevAccelBeams.23.021304} thereby potentially reducing the size and cost of such machines. Beams with separation at the plasma wave period have been produced \cite{PhysRevLett.110.065005} which can radiate coherently at terahertz frequencies. Simulations have also shown how laser-triggered ionization injection may be used to produce beams with density modulations at $\geq 50~\nano\meter$ ($h\leq 5$) \cite{xu2016nanoscale}. In this Letter, we report on how to generate a GeV-level high quality electron beam with a density modulation in the X-ray wavelength ($h\leq 1000$) range. The idea is to trigger a series of multiple downramp injections by periodically modulating the density in the downramp. 

\begin{figure}[bp]
\includegraphics[width=0.5\textwidth]{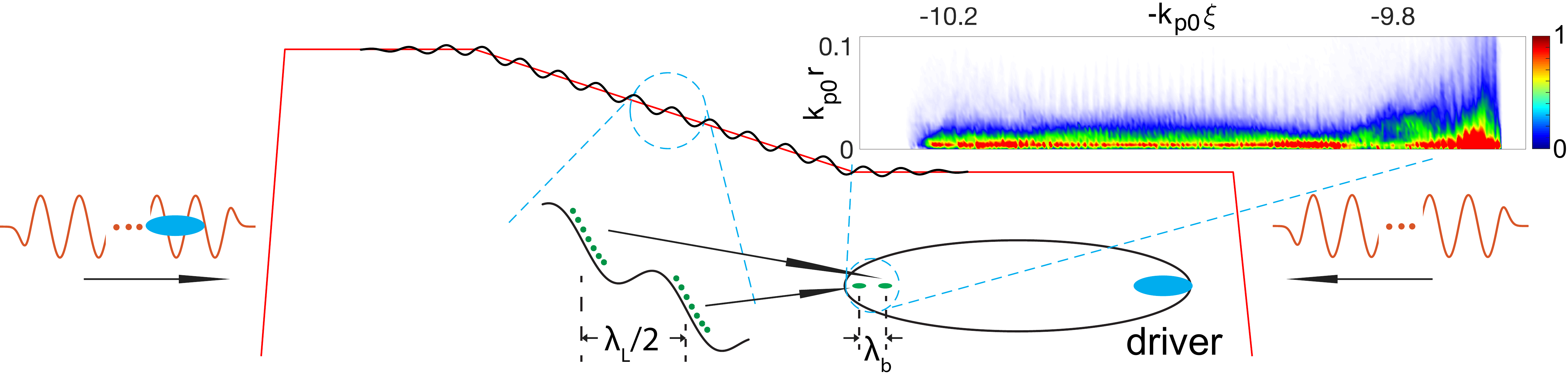}
\caption{\label{fig: concept}  The concept of bunched beam generation from density downramp injection in PBA (not to scale). A plasma with a density downramp (red line) is set at $t=0$. Two linearly polarized laser pulses propagate from the two ends of the plasma towards its center and their overlap creates a density modulation with period $\lambda_L/2$ (black line). A properly delayed electron beam driver is incident on the plasma from the left to excite a blowout wake and inject electrons only at half of the density modulation period (green dots) and they are mapped to discrete axial positions  with period $\lambda_b$ after injection. Modulated density distribution of the injected electrons from quasi-3D PIC simulations is shown in the inset. }
\end{figure}

When a short and intense driver propagates in a underdense plasma with speed $v_d\approx c$, a fully blowout plasma wave wake can be created behind it, where $c$ is the speed of light in vacuum. The phase velocity of this wake in a density gradient $n_p(z)$ can deviate from the driver's velocity due to the density dependence of the oscillation frequency \cite{PhysRevA.33.2056} as $v_\phi = \frac{v_d}{1- (\mathrm{d} \omega_p/\mathrm{d} z) \omega_p^{-1} (v_d t-z)} $, where $\omega_p(z)=\sqrt{ \frac{n_p(z)e^2}{m\epsilon_0}}$ is the local plasma frequency. A plasma with a density downramp region has  been  proposed to decrease the wake phase velocity and trap plasma electrons \cite{PhysRevE.58.R5257, PhysRevLett.86.1011, PhysRevLett.100.215004, gonsalves2011tunable, PhysRevLett.110.185006}. Recently, three-dimensional (3D) simulations found that electrons injected in the 3D blowout wake using this scheme are characterized by high current, low emittance and low slice energy spread; and theoretical analysis revealed the underlying physical mechanism that leads to the generation of high-quality electrons \cite{xu2017high}. The longitudinal dynamics of the injected electrons revealed that there was a roughly linear correlation between the longitudinal positions of the electrons before and after injection. The compression factor which is defined as the ratio of the longitudinal duration of the electrons before and after injection can be as high as several hundred \cite{xu2017high}. 



We propose to generate a microbunched high quality electron beam by superimposing a small amplitude sinusoidal density modulation on the ramp so that the density in the ramp is, $ n_p=n_{p0} \left(1- \Delta n\frac{z}{L_r} \right) + \delta \bar{n} \mathrm{sin} (k_m z)$ where $L_r$ is the length of the ramp, $\delta \bar{n}$ and $k_m$ are the amplitude and wavenumber of the modulation. The modulation only has to be on the electrons, but it can also be on the ions as well. Substituting the expression for the modulated ramp into the equation for the phase velocity of the wake gives $v_{\phi} \approx 1 + \frac{- g +  (\delta \bar{n}/n_{p0}) (k_m/k_{p0}) \mathrm{cos}({k}_m {z})}{2} k_{p0}\xi$ where $\xi = c t-z$ is the position inside the wake and $ g \equiv \frac{\Delta n/n_{p0}}{k_{p0}L_r}$ is the normalized density gradient of the ramp. This shows that  when $\frac{\delta \bar{n}}{n_{p0}} \frac{k_m}{k_{p0}} > g$, the phase velocity can oscillate between subluminal ($<c$) and superluminal ($>c$) so that injection can be turned on ($v_{\phi} < c$) and off ($v_{\phi} > c$) in the ramp. More details are provided in the supplement. After injection, the electrons are mapped to the tail of the wake sequentially, i.e., the electrons injected earlier sit in front of the electrons injected later (Fig. \ref{fig: concept}) due to the gradual elongation of the bubble in the downramp. Thus, the duration of the electrons can also be compressed significantly depending on the rate of the elongation, and the period of the charge distribution can be shrunk by several orders of magnitude, i.e., from optical frequency to X-ray frequency (Fig. \ref{fig: concept}). This bunched beam has similar properties as a beam from normal density downramp injection \cite{xu2017high}, i.e., it has high current, low emittance and low energy spread; thus it may be used to produce fully coherent, stable, hundreds of gigawatt X-rays with femtosecond duration in a short resonant undulator. 

Such electron density modulations can be realized through the ponderomotive force of  two counter-propagating linearly polarized laser pulses of the same frequency. The modulation will be at half the wavelength of lasers so injection will turn on and off with a length scale of half the optical wavelength. As illustrated in Fig. \ref{fig: concept}, two non-relativistic lasers ($a_{L}\equiv\frac{eA_L}{m_ec}\ll 1$ where $A_L$ is the vector potential of the laser) with identical frequencies propagate into the plasma from both ends and interfere inside the density ramp. This results in an electron density modulation (black line in Fig. \ref{fig: concept}) with an amplitude determined through the balance of the ponderomotive force on the electrons with the  charge separation force between the electrons and ions. 

\begin{figure}[bp]
\includegraphics[width=0.5\textwidth]{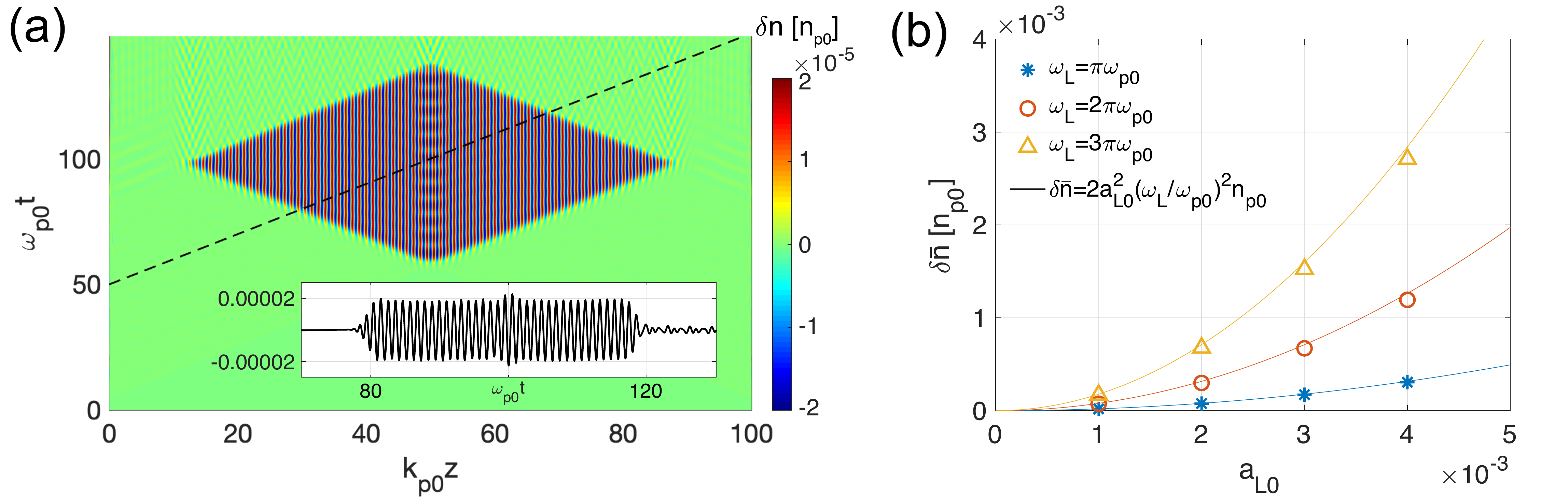}
\caption{\label{fig: ripples simulation}  (a) The perturbed density distribution induced by the colliding lasers in the $z-t$ plane from 1D simulations. The inset shows the density along the black dashed trajectory. The lasers have a rising and falling edge 10 $\omega_{p0}^{-1}$ and a plateau 60 $\omega_{p0}^{-1}$. (b) The dependence of the perturbed density on the frequency and intensity of the colliding lasers. }
\end{figure}

The amplitude of the density modulation can be estimated from $\frac{\partial^2 \delta n}{\partial t^2} + \omega_{p}^2 \delta n= \frac{e^2n_p}{2m_e^2} \frac{\partial A_L^2}{ \partial z^2}$ where $\delta n\ll n_{p}$ is assumed. If the envelop of the lasers rises slowly compared with the plasma oscillation, $\delta n$ follows the envelope of $A_L^2$. For the lasers with an approximately trapezoid shape, the perturbed density saturates at $\delta n/n_{p}\approx 2a_{L0}^2 (\omega_L/\omega_{p})^2 \mathrm{cos}(2k_Lz)$ \cite{supplement}. 1D particle-in-cell (PIC) simulations are conducted to confirm the creation of this density modulation. A cold plasma with constant density $n_{p0}$ is distributed between $z=0$ and $100k_p^{-1}$ and two lasers pulses with $\omega_L=\pi \omega_{p0}, a_{L0}=10^{-3}$ propagate towards one another from the two ends of the plasma. The plasma electron density perturbation is shown in Fig. \ref{fig: ripples simulation}(a) where a spatial sinusoidal density modulation is present in a rhombic region of the $z-t$ space. The inset shows the density seen by an observer moving at $c$ at a later time where the dashed line is its trajectory. The plasma density is modulated at wavenumber $k_m=2k_L$ and its amplitude is $\delta \bar{n}\approx 2\times 10^{-5} n_{p0}$. The frequency and intensity of the colliding lasers are scanned, and the amplitudes of the excited density modulation are summarized in Fig. \ref{fig: ripples simulation}(b). Good agreement between the simulations and the formula are obtained.

\begin{figure}[bp]
\includegraphics[width=0.5\textwidth]{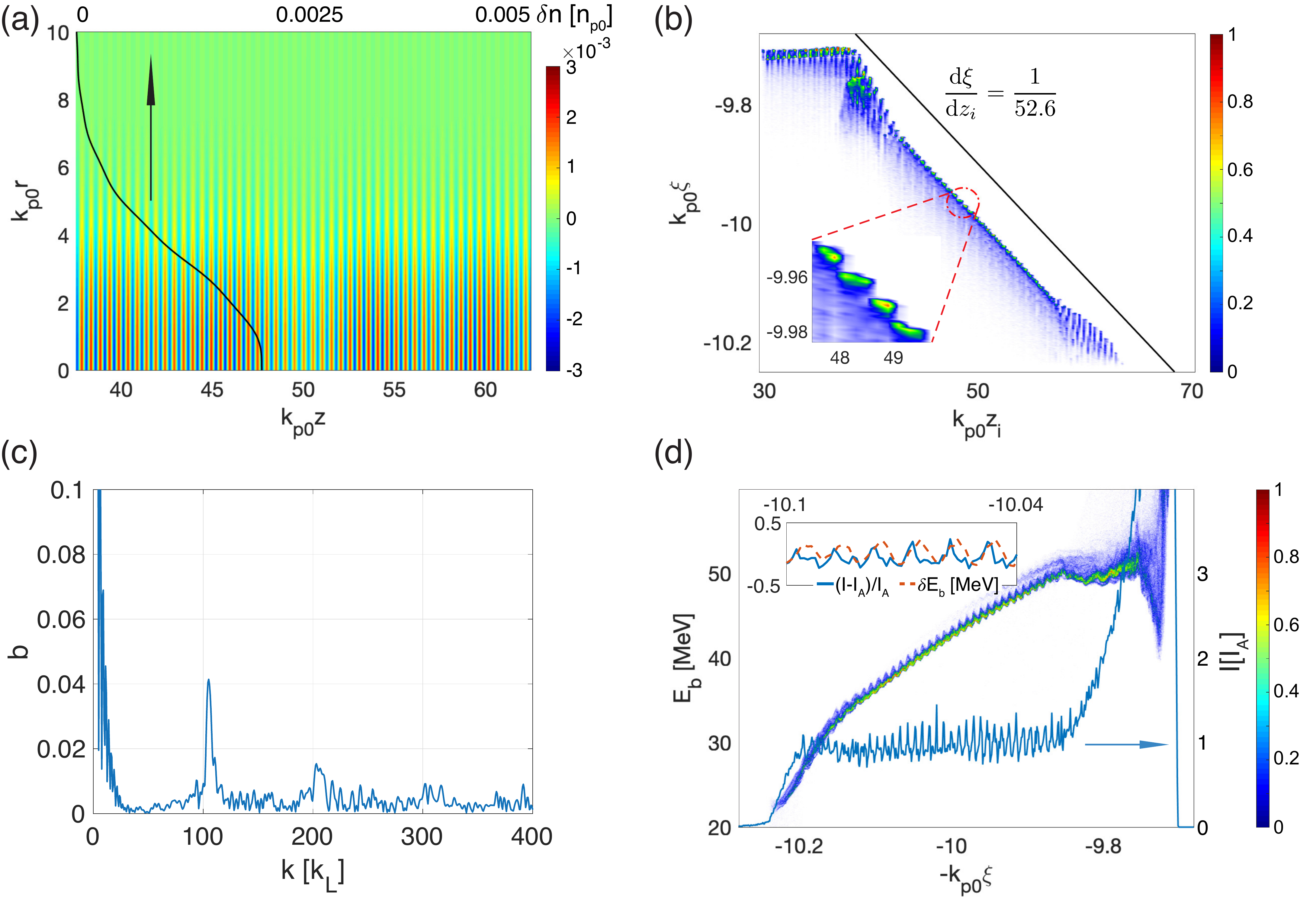}
\caption{\label{fig: self-consistent} Generation of bunched electrons from quasi-3D simulations. (a) The perturbed plasma density of the ramp region in the $r-z$ plane at $100\omega_{p0}^{-1}$. The black line is along $k_{p0}z=45.5$. (b) The relation between the initial longitudinal positions of the injected electrons and their positions inside the wake. (c) The bunching factor of the injected electrons at $145\omega_{p0}^{-1}$. (d) The longitudinal phase space of the injected electrons and their current profile (blue line) at $145\omega_{p0}^{-1}$. The inset shows the chirp-corrected energy modulation and the current. }
\end{figure}

To demonstrate the bunched beam generation, we use the quasi-3D version \cite{davidson2015implementation} of the fully relativistic PIC OSIRIS \cite{fonseca2002high}. The $m=0$ and $1$ modes (physical quantities are of the form $\mathrm{exp}(\pm i\phi)$, where $\phi$ is the azimuthal angle in the transverse plane) are included to model the linearly polarized lasers. A fixed simulation box with dimensions of $280 k_{p0}^{-1} \times  12 k_{p0}^{-1}$ and with $143360 \times 3072$ grids along the $z$ and $r$ directions, respectively. This corresponds to a grid size along the $z$ of $\frac{1}{512}k_{p0}^{-1}$ which is needed to resolve the short scales of the injected and trapped electrons. Each cell contains 8 macro-particles to represent the beam driver and the plasma electrons respectively. A Maxwell solver which can model relativistic particles with high fidelity is used \cite{xu2020numerical}. As illustrated in Fig. \ref{fig: concept}(a), a plasma starts from $z=0$ with a density $1.1 n_{p0}$ that drops linearly to $n_{p0}$ from $37.5k_p^{-1}$ to $62.5k_p^{-1}$ and extends to $100k_{p0}^{-1}$ with $n_{p0}$. The normalized density gradient of this ramp is $g=0.004$. Two laser pulses, polarized along $\hat x$ with $a_{L0}=0.005$ and $\omega_L= 2\pi \omega_{p0}$ are sent from both ends of the plasma and propagate toward one another. The lasers have the same longitudinal profile as shown in Fig. 2, and a Gaussian transverse profile with a $6k_{p0}^{-1}$ spot size at focus and a focal plane of $50k_{p0}^{-1}$. A plasma density modulation with wavelength $\lambda_{m}\approx \frac{\lambda_L}{2}$ is observed around the ramp region as shown in Fig. \ref{fig: self-consistent}(a). The black line shows the modulation amplitude at $45.5k_{p0}^{-1}$ which varies along the transverse directions due to the Gaussian transverse distribution of the lasers. The blowout radius of the wake is $\sim4k_{p0}^{-1}$, so the modulations are relatively constant in the region of physical interest.

The electron beam driver has an energy of $2\giga\electronvolt$, a peak current of $34\kilo\ampere$, a spot size of $k_{p0}\sigma_r=0.5$, and a duration of $k_{p0}\sigma_z=0.7$. Plasma electrons are completely expelled by the electric field of the driver and some of them are pulled back toward the axis by the immobile ions and form a high-density sheath \cite{PhysRevLett.96.165002}. When propagating through a region of  decreasing density, the wake expands, and some sheath electrons are injected at the tail end of the wake. These electrons are longitudinally locked with the driver since they are both highly relativistic. We track these injected electrons and show their initial positions $z_i$ and their relative positions inside the wake $\xi$ after injection in Fig. \ref{fig: self-consistent}(b). It is obvious that the injection position is modulated at the plasma density modulation wavelength ($0.5k_{p0}^{-1}$) as shown in the enlarged inset. There is a mapping between the initial and final position of an electron after injection \cite{xu2017high}, i.e., $\frac{\mathrm{d} \xi}{\mathrm{d} z_i} \approx 4.5g $ where the wake wavelength $k_p(z)\lambda_{wake}\approx 9$  observed from simulations is used. Thus, the macro length and micro modulation period are compressed by a factor of $\frac{1}{4.5g} \approx 55.6 $ for $g=0.004$. A black line with $\frac{\mathrm{d}\xi}{\mathrm{d}z_i }= \frac{1}{52.6}$ is shown for reference. Taking into account that $k_m=2k_L$,   the harmonic number, defined as the ratio of the colliding laser period and the density modulation period of the injected electrons,  is $h=2\left( \frac{\mathrm{d} \xi}{\mathrm{d} z_i} \right) ^{-1}\approx 0.44g^{-1}$, i.e., $h\approx 110$ for $g=0.004$. We note there is also some discrete injection before and near the end of the ramp, however the electrons injected from these regions are not bunched, as evidenced by constant value of $\xi$ even though $z$ is changing, due to the lack of the above linear mapping which is only present in the ramp.

A bunching factor $b(k)=|\sum_{j=1}^N \mathrm{exp}(i k z_j)|/N$ is usually used to quantify the modulation, where $N$ is the total number of the electrons. As shown in Fig. \ref{fig: self-consistent}(c), the bunching factor of the injected beam reaches its maximum at $k=hk_L \approx 105 k_{L}$ with $b\approx 0.042$ and the second (third) harmonic appears at $k\approx 210 k_{L} (315 k_L)$ with $b\approx0.015 (0.009)$. We note that only electrons with $k_{p0}\xi\leq -9.85 $ are used to calculate the bunching factor. 

The longitudinal phase space of the injected electrons and its current profile at $t=145\omega_{p0}^{-1}$ are shown in Fig. \ref{fig: self-consistent}(d), where the bunched structure is seen clearly. The beam has a positive energy chirp ($\frac{\mathrm{d}E_b}{\mathrm{d} z} >0$) since the electrons at the head are injected at earlier times and are thus accelerated over a longer distance \cite{xu2017high}. A sinusoidal energy modulation with an amplitude $\sim \mathrm{MeV}$ is present in the longitudinal phase space. This is caused by the axial space-charge interactions between the bunched electrons while they are at low energies because the axial electric field of relativistically moving electrons decreases rapidly with their energy ($E_{SC,z} \propto \gamma^{-2}$). The chirp-corrected average energy modulation and the current profile of the beam are shown in the inset of Fig. \ref{fig: self-consistent}(d). We can see the energy modulation is approximately $\frac{\pi}{2}$ ahead of the currents (density) modulation in phase which confirms that the space charge forces from the current modulation are the cause of the sinusoidal energy modulation. The degradation of the bunched structure induced by this energy modulation when the beam is boosted to higher energy is negligible. The slippage induced by the energy modulation is $\delta s\approx \int \frac{\delta \gamma}{(\gamma+E_z z)^3}dz \leq \frac{1}{2E_z}\frac{\delta \gamma}{\gamma_i^2} \approx 10^{-4} k_{p0}^{-1}$, which is much less than the bunching wavelength $0.01k_{p0}^{-1}$ when $E_z=2\frac{mc\omega_{p0}}{e}, \delta \gamma=2,$ and $\gamma_i=80$ are substituted into the formula.

This beam is characterized by a slice energy spread of $\sim0.4~\mathrm{MeV}$, an average current of $\sim 17 \kilo\ampere$,  and a normalized emittance of $\sim 0.006k_{p0}^{-1}$, which is suitable to drive an XFEL. However, there is a large energy chirp [$\sim 50\mathrm{MeV}/ (k_{p0}\xi)$] formed during the injection process. However, this chirp can be mitigated by an inverse chirp naturally imposed by the acceleration gradient inside the wake \cite{PhysRevLett.96.165002} (electrons injected first reside farther forward in the wake where the acceleration gradient is smaller). Thus, there is an optimized acceleration distance where the beam achieves a flatter longitudinal phase space \cite{xu2017high} that minimizes the projected energy spread of the macrobunch.

\begin{figure}[bp]
\includegraphics[width=0.5\textwidth]{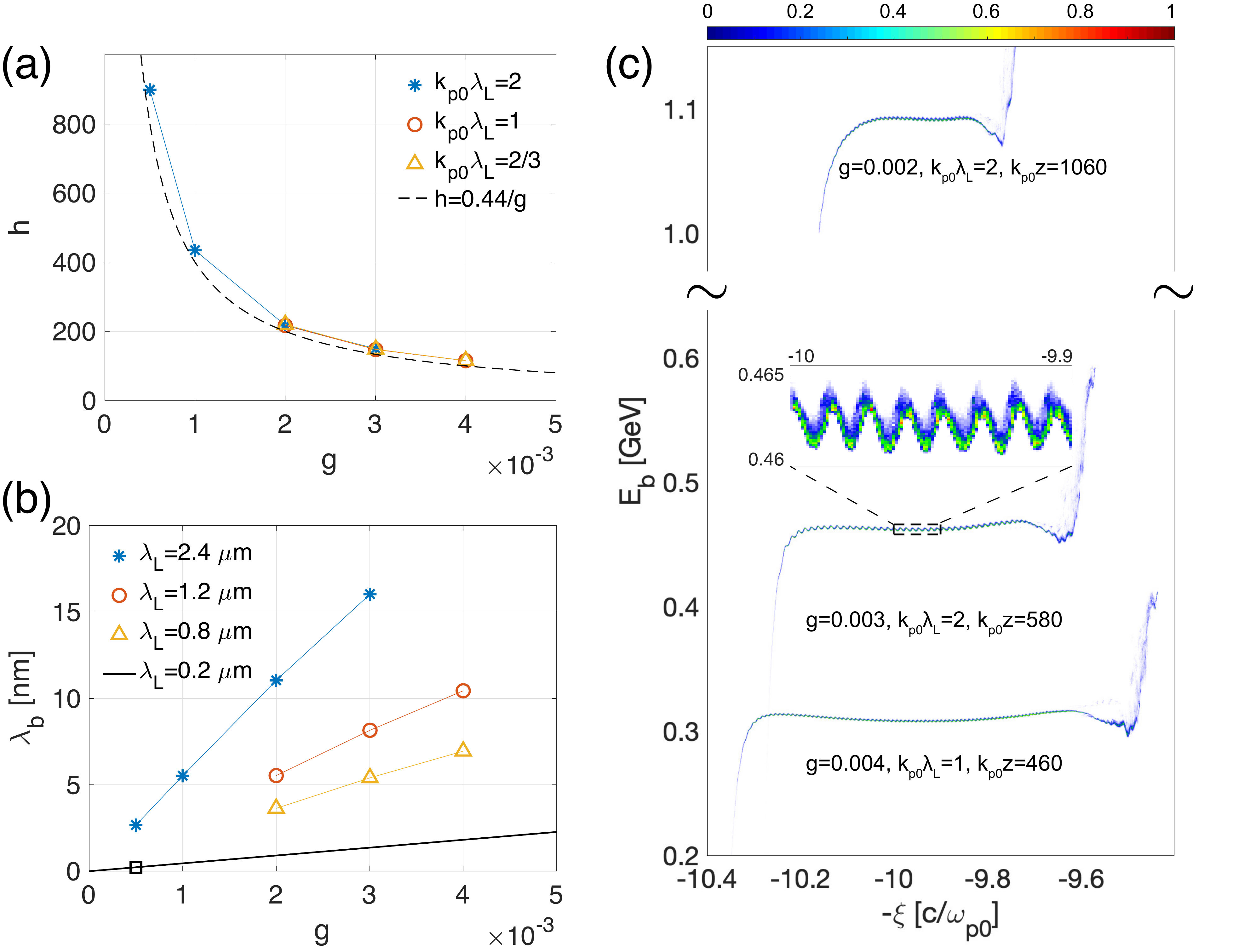}
\caption{\label{fig: scan}  (a) The dependence of harmonic number $h$ on the normalized density gradient $g$ under different colliding lasers. (b) The modulation wavelength when different ramps and lasers are used where $n_{p0}=1.97 \times 10^{19}\centi\meter^{-3}$. (c) The longitudinal phase space of the injected electrons at their optimized acceleration distance. The inset shows the sinusoidal energy modulation of case: $g=0.003, k_{p0}\lambda_L=2$. }
\end{figure}

In order to understand how to choose plasmas parameters to optimize the process, we performed a wide parameter scan using quasi-3D OSIRIS with only the $m=0$ mode where a density modulation with $\delta \bar{n}=10^{-3}n_{p0}$ is initialized on top of a plasma downramp.  Since we are not including self-consistently lasers to create the modulation, we use a moving window with dimensions $11.25 k_{p0}^{-1} \times 12 k_{p0}^{-1}$. More details for the parameters of the simulations can be found in the supplement. To isolate the physics, we also employ a non-evolving beam driver to model the injection and acceleration. The dependence of the harmonic number $h$ on the gradient of the ramp $g$ is shown in Fig. \ref{fig: scan}(a) (where the theory curve $h= 0.44/g$ is also plotted). Clearly $h$ only depends on the gradient of the ramp $g$, not on the modulation period. When using a ramp with $g=5\times10^{-4}~(\Delta n=0.05n_{p0}, k_{p0}L=100)$, the modulation frequency of the injected electrons is as high as 900 times of the laser frequency, i.e., $h\approx 900$, which is one order of magnitude higher than the number from staged HGHG ($h\leq 60 $ \cite{allaria2013two}) and EEHG ($h\leq 101$ \cite{ribivc2019coherent}). Normalized units are used in the above results where the modulation frequency is scaled to the plasma frequency. We can choose a normalized density to make connections with possible near-term experiments. Fig. \ref{fig: scan}(b) shows the bunching wavelength vs g for various $\lambda_L$ for $n_{p0}=1.97\times10^{19}~\centi\meter^{-3}$ (assuming that the density wavelength is half  $\lambda_L$) and it can be seen that modulations at several nm are achieved. In principle, beams with shorter modulated wavelength can be injected if the ramp density gradient is smaller or the density modulation wavelength of the ramp is shorter, which might be the case for shorter wavelength lasers. The predicted results when $\lambda_L=0.2~\micro\meter $ is shown in black line in Fig. \ref{fig: scan}(b) where 0.23 nm bunching wavelength is achieved when $g=5\times 10^{-4}$.

In Fig. \ref{fig: scan}(c), we show that that when the injected electrons are accelerated to an optimized distance \cite{xu2017high}  the projected energy spread of the beam core is less than 1 $\mathrm{MeV}$. Since the initial energy chirp induced during the injection process is  inversely proportional to the gradient of the ramp, i.e., $\frac{\mathrm{d}E_b}{\mathrm{d}z} \propto g^{-1}$, a long acceleration distance is required to remove it if $g$ is small. This will also lead to a higher energy beam. Beam loading \cite{PhysRevLett.101.145002} which is included in the simulations can also play a role during the  removal of the chirp. 

As noted above the  longitudinal space charge can create a $\sim\mathrm{MeV}$ sinusoidal energy modulation along the beam as shown in the inset of Fig. \ref{fig: scan}(c). By letting the beam go through a chicane where the electrons' axial location changes according to their energy, this energy modulation can be utilized to enhance the bunching factor at the $hk_L$ and their harmonics as in a HGHG \cite{yu1991generation}. For example, a small chicane with $r_{56}=4.1 \mathrm{nm}/\mathrm{MeV}$ can increase the bunching factor of the beam from $g=0.003, k_{p0}\lambda_L=2$ at the fundamental frequency from $0.05$ to $0.24$ and at the third harmonic from $0.02$ to $0.08$.

In order to transport the beams out of the  plasma into an undulator without emittance degradation a plasma matching section with gradually varying density is needed \cite{PhysRevSTAB.17.054402, PhysRevSTAB.18.041302, xu2016matching} to match the beam.  The matched beam can then travel through a resonant undulator to generate  fully coherent radiation. Due to the high current ($\sim 17 \kilo\ampere$) and low emittance ( $\sim 0.01 k_{p0}^{-1}\approx 12\nano\meter $) of these beams, hundreds of GW of radiation can be emitted in a short beam-undulator interaction distance. Taking the 1.09 GeV injected beam from $g=0.002, \lambda_L=0.8~\micro\meter$ case as an example \cite{explanation}, 3.6 $\nano\meter$, fully coherent radiation with a stable power of $\sim 0.3$ TW could be emitted from a meter long resonant undulator with $\lambda_U=1\centi\meter, K=2.1$ \cite{xie1995design}. 


In conclusion, we have proposed a novel approach that uses a PWFA and density modulated downramp to produce ultra-bright and high-quality electron bunches  with a current modulation at X-ray wavelengths.  We have presented fully self-consistent PIC simulations that show that the ponderomotive force of two counter propagating lasers that overlap in the downramp can produce a sufficient density modulation with a wavelength half of the laser wavelength. The current of the injected beam is modulated  at a wavelength $O(1000)$ times smaller than the modulation wavelength in the downramp. 
\begin{acknowledgments}
Work supported by the U.S. Department of Energy under contract number DE-AC02-76SF00515, No. DE-SC0010064, and SciDAC FNAL subcontract 644405, and NSF Grants Nos. 1734315 and ACI-1339893. The simulations were performed on the UCLA Hoffman 2 and Dawson 2 Clusters, and the resources of the National Energy Research Scientific Computing Center.
\end{acknowledgments}

\bibliography{refs_xinlu}

\begin{thebibliography}{43}
\bibitem[{exp()}]{explanation}
\bibinfo{note}{Although the longitudinal phase space shown in Fig. 4(c) is
  using $g=0.002, \lambda_L=2.4\micro\meter$, we assume the beam when $g=0.002,
  \lambda_L=0.8\micro\meter$ has a similar distribution since the fine
  structures of the beam affect the acceleration little.}
\end{thebibliography}

\end{document}